\documentclass[lettersize,journal]{IEEEtran}
\usepackage{amsmath}
\usepackage{amssymb}
\usepackage{graphicx}
\usepackage{comment}
\usepackage{smartdiagram}
\usepackage{soul}
\usepackage[caption=false,font=normalsize,labelfont=sf,textfont=sf]{subfig}
\newcommand{\ac}[1]{\textcolor{black}{#1}}
\newcommand{\bc}[1]{\textcolor{black}{#1}}
\newcommand{\an}[1]{\textcolor{black}{#1}}

\begin{document}

\title{Optimizing Smart Helper Placement for Enhanced Cache Efficiency in F-RANs}

\author{\IEEEauthorblockN{Hesameddin~Mokhtarzadeh, \IEEEmembership{Graduate Student Member,~IEEE,} Mohammed~Saif, ~\IEEEmembership{Member,~IEEE,} \\Md. Jahangir Hossain, \IEEEmembership{Senior~Member,~IEEE,} Julian Cheng, \IEEEmembership{Fellow,~IEEE} }\\

\IEEEcompsocitemizethanks{\IEEEcompsocthanksitem Hesameddin~Mokhtarzadeh, Md. Jahangir Hossain, and Julian Cheng are with the School of Engineering, The University of British Columbia, Kelowna, BC V1V 1V7, Canada (e-mails: hesameddin.mokhtarzadeh@ubc.ca, jahangir.hossain@ubc.ca, julian.cheng@ubc.ca).

Mohammed Saif is with the Department of Electrical and Computer Engineering, University of Toronto, Toronto, ON M5S, Canada (e-mail: mohammed.saif@utoronto.ca).
}	
}

\maketitle

\begin{abstract}
\bc{Smart helpers (SHs) have been proposed to improve content delivery delays and alleviate high fronthaul loads in fog radio access networks (F-RANs). They offer an alternative to deploying additional \an{enhanced remote radio heads (RRHs)}, which are often infeasible due to site constraints.} The optimal placement of SHs can significantly increase the number of users they serve which leads to enhanced cache efficiency and improved content delivery delay.
In this letter, we optimize SH placement within an F-RAN to maximize the cache hit rate and further reduce the content delivery latency. 
We model the \an{SH cache hit rate} as a function of outage probability and user density distribution. 
We develop a function to estimate user density distribution leveraging the radial basis functions (RBFs) method and optimize SH placement utilizing the particle swarm optimization (PSO) algorithm. \an{Our} numerical results confirm the effectiveness of the proposed approach in maximizing the \an{SH cache hit rate}, thereby improving delivery delays and fronthaul loads of the network.
\end{abstract}


\section{Introduction}
\IEEEPARstart{F}{og}-radio access networks (F-RANs) have become a promising solution for ultra-reliable low-latency services by upgrading \ac{remote radio heads (RRHs)} to enhanced RRHs (eRRHs), \ac{bringing} caching and signal processing closer to users 
\cite{7558153}. 
The demand for improved Quality of Service (QoS) pushes for more eRRH installations in F-RANs. However, densifying these networks by increasing the number of eRRHs presents significant challenges, \ac{including} space limitations at potential sites and \ac{the cost and complexity} of establishing additional fronthaul links, particularly in densely populated urban areas.


Cache-enabled device-to-device (CE-D2D) communications have emerged as a viable solution to enhance QoS in F-RANs by allowing devices to cache popular content and serve nearby users, reducing the dependency on additional eRRHs \cite{6928445,D2D}. \bc{Specifically, the congestion on the eRRHs and the heavy loads on fronthaul links could be alleviated by sharing the content between the users \cite{RW4,RW5}.}
However, despite their potential, CE-D2D communications have \ac{some limitations} due to privacy and power usage concerns. 
\ac{On the other hand,} CE-relays have been \ac{also} explored as a solution to cache content and facilitate communication between eRRHs and users to improve QoS~\cite{CERelay1}. \bc{These relays cache popular content while operating as an intermediary between users and eRRHs and serve \an{users} with their requests in the subsequent time frames \cite{CERelay3}.}
Nevertheless, CE-relays face inefficiencies in cache updates when used infrequently, and when used frequently, they introduce delays and increase fronthaul load due to the added communication between eRRHs, CE-relays, and the MBS to retrieve the necessary data \cite{CERelay2}.

\bc{As an alternative to CE-D2D and CE-relay communications,} \cite{10818447} proposed the concept of smart helpers (SHs) to enhance the QoS without expanding eRRHs and fronthaul links, while \ac{addressing} CE-D2D security concerns and improving cache update and delivery delays compared to CE-relays. 
\ac{In the proposed SH-aided F-RAN, SHs are cost-effective elements that do not need to establish extra fronthaul links. They can listen to the communications between the eRRHs and the users happening around them, cache the popular content smartly, and serve users with their requests. To optimize the performance of SHs, we tackled the problem of minimizing delays by optimizing both resource allocation and cache management for SHs, when they are positioned at random locations within the network.} However, optimizing the SH placement in a network with a non-uniform distribution of users has a direct impact on the cache hit rate of SHs due to their service coverage. 

In this letter, we optimize the placement of SHs in F-RANs to maximize cache hit rates and reduce latency. We model the SH hit rate, defined as the number of users under the SH's coverage, as a function of outage probability and user density distribution. While user density is inherently random and varies over time, previous studies \cite{7248881,6757900} have shown that it exhibits long-range temporal dependence and region-specific characteristics. We estimate user density through eRRH activity history via the radial basis function (RBF) algorithm \cite{7083041}. We then formulate an optimization problem for SH placement, which is non-convex and analytically intractable. We solve the problem using the particle swarm optimization (PSO). Numerical results demonstrate the effectiveness of our approach, which can also guide network planning by identifying regions where additional eRRHs or relays would improve coverage and performance.

\section{System Model}
\label{sec:system_model}
\subsection{Spatial Distribution of eRRHs and Users}
\ac{Consider an F-RAN network, where $R$ single-antenna eRRHs and a single-antenna SH serve \an{$U$} single-antenna users within a rectangular area~$\mathcal{A}$.} \bc{We denote the set of eRRHs, the SH, and the set of users by $\mathcal{R}$, $\mathcal{S}$, and $\mathcal{U}$, respectively.} The eRRHs are spatially distributed following a \ac{poisson point process (PPP)} with an intensity $\lambda_\mathcal{R}$. \ac{The set of eRRHs are represented} as $\mathcal{R} = \{l_r; r \in \mathbb{N}\},$ where $l_r = (x_r, y_r)$ denotes the location of \ac{the $r$th eRRH}.
\ac{We assume that} the spatial density of users within a specific area remains stationary over a relatively long time frame. This means that while individual user locations may change randomly, the overall statistical properties of user density are consistent over time, influenced by stable factors such as population distribution and geographical constraints.
\ac{Following \cite{8108153, 7733098}, we model the user density distribution using PPP}. Additionally, a user-eRRH correlated \ac{poisson cluster process} (PCP) has been considered in \cite{9860085, 8023448}. However, in dense and crowded urban areas, \an{spatial distribution of users} often becomes independent of eRRHs due to dynamic environmental changes, such as new building constructions and company launches. Hence, we adopt an independent \ac{PCP} to model the user density distribution. Specifically, we assume that users form clusters representing buildings and groups of people within the area. The complete set of PCP users is denoted by 
$\mathcal{C} = \bigcup_{c} \left(l_c + \mathcal{U}_c \right)$,
where $l_c$ indicates the location of cluster head $c$, and $\mathcal{U}_c = \{l^{\mathcal{C}}_{c,u}; u \in \mathbb{N}\}$ represents the users in cluster $c$, \ac{where} $l^{\mathcal{C}}_{c,u}$ denotes the location of user $u$ within cluster $c$. Within each cluster, $N_c$ users are scattered according to a circularly symmetric normal distribution with variance $\sigma^2_{\mathcal{U}}$ around the cluster head. 
To account for pedestrians and individually scattered users, we introduce an additional set of users, $\mathcal{P} = \{l^{\mathcal{P}}_u; u \in \mathbb{N}\}$, which is distributed according to a PPP with an intensity $\lambda_\mathcal{U}$. Here, $l^{\mathcal{P}}_u$ represents the location of user $u$ in $\mathcal{P}$.
Finally, we combine both sets of users to represent the total user population as
$\mathcal{U}^{T} = \left\{ \mathcal{C}, \mathcal{P}\right\}.$


\subsection{User Scheduling and Outage Probability of the SH}
\ac{The channel between each pair of nodes (i.e., users and eRRHs/SH) is modeled using the Rayleigh fading channel model, which is widely used in environments where there are many scatters and no line-of-sight path is dominant \cite{6171996}}. The channel between a typical user \ac{that is located at} $l_{\mathcal{U}} = (x,y)$ and the eRRHs and the SH is given by 
\begin{align} \label{channel}
h_{\mathcal{U},\zeta} = \sqrt{\frac{\beta_0}{\lVert l_u - l_{\zeta} \rVert^{\alpha}}} \tilde{h}_{u,\zeta},
\end{align}
where $\zeta \in \{\mathcal{R},\mathcal{S}\}$ represents one of the eRRHs or the SH, $ \beta_0 $ is the path loss at the reference distance $d=1m$, $ \lVert l_r - l_{\zeta} \rVert $ is the distance between the user and $\zeta$, $ \alpha $ is the path loss exponent, and $\tilde{h}_{u,\zeta} \sim \mathcal{CN}(0,1)$ represents the Rayleigh fading component. 

We assume that the eRRHs and the SH serve the users using different sets of frequency-orthogonal resource blocks. This orthogonality ensures no frequency overlap between the resources allocated to the users by the associated eRRH or the SH. However, users experience interference from other eRRHs and the SH using the same set of resource blocks. 
Hence, the received signal at the intended user $u$ is given by
\begin{align}
y_{u} = \sqrt{P_T} h_{u,\zeta} a_{u,\zeta} + \sum_{\underset{j \neq \zeta}{j\in \mathcal{R}_\zeta} } \sqrt{P_T} h_{u,\zeta} a_{u,j} + n_u,
\end{align}
where $a_{u,\zeta} \in \{0,1\}$ represents the assignment of user $u$ to eRRH $\zeta$, $n_u \sim \mathcal{CN}(0,N_0)$ is the received noise with $N_0$ \ac{as the} noise power, $\mathcal{R}_\zeta \subset \mathcal{R} $ is the set of eRRHs utilizing the same frequency set as $\zeta$, and $P_T$ is the transmit power from $\zeta$. 
Thus, the signal-to-interference-plus-noise ratio (SINR) \ac{of user $u$ that is associated with $\zeta$} is given by
\begin{align}
\gamma_{u,\zeta} = \frac{P_T |h_{u,\zeta}|^2}{\sum_{\underset{j \neq \zeta}{j\in \mathcal{R}_\zeta} } P_T |h_{u,j}|^2 + N_0}.
\end{align}
Similar to \cite{6042301}, we assume that the noise power at the users' receivers is negligible compared to the interference signal power they receive. Hence, \an{the users are assigned to the eRRH based on} the maximum signal-to-interference (SIR) value. We denote $n_r$ as the number of assigned users to the $r$th eRRH, which is given by 
$n_r = \sum_{u \in \mathcal{U}^{T}} \mathbb{I}(\arg \max_{r'\in \mathcal{R}} \gamma_{u,r} =~r), 
$ where $\mathbb{I}(.)$ is the indicator function. 
In addition, the average outage probability of serving a typical user in location $l_{\mathcal{U}} = (x,y)$ by the SH at the location $l_{\mathcal{S}} =(x_{\mathcal{S}},y_{\mathcal{S}})$ \ac{can be} obtained as 
\begin{align}
P_{out}(x,y|l_{\mathcal{S}}) = \mathbb{P}\{ \frac{P_T |h_{\mathcal{U},\mathcal{S}}|^2}{\sum_{r\in \mathcal{R}_\zeta} P_T |h_{\mathcal{U},r}|^2} \leq \bar{\gamma}|x,y\}, 
\end{align}
where $\bar{\gamma}$ is the minimum required SIR value \bc{between the SH and a typical user at the location of $(x,y)$} to make a reliable connection. \ac{Considering (\ref{channel}),} we have 
\begin{align}
P_{out}(x,y|l_{\mathcal{S}}) = \mathbb{P} \{|\tilde{h}_{\mathcal{U},\mathcal{S}}|^2 \leq 
\bar{\gamma} \sum_{r\in \mathcal{R}_\zeta}   \left( \frac{\lVert l_{\mathcal{S}} - l_{\mathcal{U}} \rVert}{\lVert l_r - l_{\mathcal{U}} \rVert} \right)^{\alpha} |\tilde{h}_{\mathcal{U},r}|^2 \},
\end{align}
where $|\tilde{h}_{\mathcal{U},\mathcal{S}}|^2$ follows a $\chi$-square distribution with two degrees of freedom that is equivalent to an exponential distribution with a mean $1$. Hence, the \ac{outage probability} for a single user \ac{at} the location $(x,y)$ and the SH at the location $(x_{\mathcal{S}},y_{\mathcal{S}})$ is given by
\begin{align}
P_{out}(x,y|l_{\mathcal{S}}) = 1- \prod_{r\in \mathcal{R}_\zeta} \mathbb{E} \{e^{-\bar{\gamma}  \left( \frac{\lVert l_{\mathcal{S}} - l_{\mathcal{U}} \rVert}{\lVert l_r - l_{\mathcal{U}} \rVert} \right)^{\alpha} |\tilde{h}_{u,r}|^2}\}.
\end{align}
Since $|\tilde{h}_{\mathcal{U},r}|^2$ is also exponentially distributed, the expectation simplifies the expression further as follows \ac{\cite{6524460}}
\begin{align}
P_{out}(x,y|l_{\mathcal{S}})= 1- \prod_{r\in \mathcal{R}_\zeta} \frac{1}{1+\bar{\gamma}  \left( \frac{\lVert l_{\mathcal{S}} - l_{\mathcal{U}} \rVert}{\lVert l_r - l_{\mathcal{U}} \rVert} \right)^{\alpha}}.
\end{align}
Now, given the \an{spatial density distribution $D(x,y)$}, the average number of users under the coverage area of the SH at the location of $l_{\mathcal{S}}$ \ac{is obtained by} 
\begin{align} \label{ns}
n_{\mathcal{S}}(l_{\mathcal{S}}) = \int_{\mathcal{A}} \prod_{r\in \mathcal{R}_\zeta} \frac{1}{1+\bar{\gamma}  \left( \frac{\lVert l_{\mathcal{S}} - l_{\mathcal{U}} \rVert}{\lVert l_r - l_{\mathcal{U}} \rVert} \right)^{\alpha}} \times D(x,y)~dx~dy.
\end{align}
\section{Problem Formulation}
\label{sec:problem_formulation}
\begin{figure}
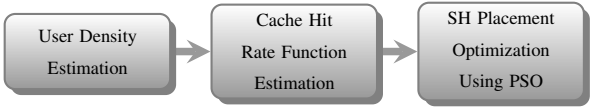
 
\centering \smartdiagramset{uniform color list=white!60!black for 6 items, module minimum width=2.2cm, font=\fontsize{7pt}{12pt}\selectfont, back arrow disabled=true}
\smartdiagram[flow diagram:horizontal]{User Density Estimation, Cache Hit Rate Function Estimation, SH Placement Optimization Using PSO}
\caption{Diagram of the proposed solution for optimizing the SH \ac{placement}.}\label{fig1}
\end{figure}

In general, the average number of users served by the SH \an{that is located at  $(x_{\mathcal{S}}, y_{\mathcal{S}})$} is given by
\begin{align}
N^{cache}_{\mathcal{S}}(x_{\mathcal{S}}, y_{\mathcal{S}}) = n_{\mathcal{S}}(x_{\mathcal{S}}, y_{\mathcal{S}}) \times P_c,
\end{align}
where $P_c = \sum_{f=1}^F P_p  z_f c_f$ is the cache hit success probability of the SH, $c_f$ represents the optimized cache resource, $P_p$ is the probability of requesting popular file segments, and $z_f$ is the probability of requesting the $f$th popular file segment. Due to the independence of $P_c$ from the location of the SH, we assume $P_c$ is optimized using \ac{the} algorithm proposed in \cite{10818447}, and focus on maximizing the cache hit rate by optimizing the location of the SH. \an{Given optimized $P_c$, the SH  placement optimization problem that aims to maximize the average number of users under SH coverage while ensuring the area boundary constraints is formulated as follows}
\begin{align} \label{Problem}
	&\max_{x_{\mathcal{S}},y_{\mathcal{S}}} ~~~~ n_{\mathcal{S}}(x_{\mathcal{S}},y_{\mathcal{S}})\nonumber\\
	&~~~\textrm{s.~t.~:}   
    \begin{array}{clc}
    &(x_{\mathcal{S}},y_{\mathcal{S}}) \in \mathcal{A}.
    \end{array} 
\end{align}
Obtaining the optimal solution to this problem is analytically challenging due to a couple of reasons. First, the \an{spatial density distribution function $ D(x,y)$} is not known and needs to be estimated. Second, the product term $\prod_{r \in \mathcal{R}_\zeta} \frac{1}{1+\bar{\gamma} \left( \frac{\lVert l_{\mathcal{S}} - l_{\mathcal{U}} \rVert}{\lVert l_r - l_{\mathcal{U}} \rVert} \right)^{\alpha}}$ incorporates non-linearity from the exponent $\alpha$ and the Euclidean distances. \ac{Hence, the analytical solution is intractable.} To tackle this \an{challenge}, we \ac{first} develop an algorithm to estimate the user density distribution using RBF algorithm \ac{as in Section \ref{sec:kde}}. Then, \ac{using the estimated user density distribution}, we estimate the hit rate function and present an algorithm for optimizing the \an{SH placement} using the \ac{particle swarm optimization (PSO)} algorithm as \ac{presented in Section~\ref{sec:pso}}. \ac{Figure~\ref{fig1}} represents the diagram of the proposed approach for optimizing the \an{SH placement}.
\section{Proposed solution: estimation of user density and SH placement optimization}

\subsection{Estimation of User Density}
\label{sec:kde}
In our case, user assignment depends on the distance between eRRHs and users since the users are assigned to the eRRH with \an{the maximum} SIR. To model the user density distribution, RBFs are particularly effective, as they account for distance variations and provide smooth interpolation of irregularly distributed data \ac{\cite{7083041}}. Therefore, we model the user density function $D(x,y)$ using RBFs centred at the eRRH locations.
\ac{However, the non-uniform distribution of eRRHs creates challenges in accurately assessing user density distribution based solely on the number of users associated with each eRRH \cite{7248881}. To address this non-uniformity, we normalize the number of assigned users to the eRRHs to the Voronoi area of each eRRH.} 
Let $\mathcal{A}_r \subset \mathcal{A} $ denote the Voronoi area of \ac{the $r$th eRRH}. The normalized number of users associated with \ac{the $r$th eRRH} is defined as $\bar{n}_r = n_r / \text{area}(\mathcal{A}_r)$.
Given the location $l_r$ of the $r$th eRRH and its normalized number of users $\bar{n}_r$, we estimate the weighted \ac{density distribution of the users}  as
\begin{align}
\tilde{D}(x,y) = \sum_{r=1}^{R} w_r \varphi(x_r,y_r|l_r),
\end{align}
where $\varphi(\cdot)$ is the RBF and $w_r$ is the interpolation weight of eRRH $r$. Here we consider the linear function as the RBFs of eRRHs  
$\varphi(x,y|l_r) = \lVert l_{r} - l_{\mathcal{U}} \rVert$,  where $l_{\mathcal{U}} =(x,y)$ and $l_{r} =(x_r,y_r)$.
The weights \( w_r \) are determined by solving the interpolation system formed by enforcing the condition that \( \tilde{D}(x_r, y_r) = \bar{n}_r \) at each eRRH location. This leads to a system of linear equations as
$\mathbf{w} = \Phi^{-1} \mathbf{\bar{n}},$ where $ \Phi $ is the matrix with elements $ \Phi_{ij} = \varphi(\lVert l_i - l_j \rVert) $, and $ \mathbf{\bar{n}} = [\bar{n}_1, \bar{n}_2, \dots, \bar{n}_R]^T $ is the vector of normalized user counts at each eRRH. By solving this system, we obtain the weights $ w_r $, which are then used to interpolate the user density across the network. 
To ensure that the estimated user density $\tilde{D}(x,y)$ accurately reflects the actual total number of users, we normalize it by multiplying \ac{$\tilde{D}(x,y)$} by the factor $\frac{U}{\int_{\mathcal{A}} \tilde{D}(x,y) \, dx \, dy}$. 
\subsection{Optimizing SH Location Using PSO}
\label{sec:pso}
\bc{Given user density distribution estimate, the number of users covered by the SH at location $l_{\mathcal{S}}$ can be determined using Equation (\ref{ns}).} Now, we employ \ac{PSO} to optimize the location of the SH. The objective is to maximize the fitness function $n_{\mathcal{S}}(l_{\mathcal{S}})$, where $l_{\mathcal{S}}$ represents the SH location.
The algorithm begins by randomly initializing a swarm of particles, $l_{\mathcal{S}}^i(0)$ for $i \in \{1, \dots, M\}$, with each particle representing a potential location for the SH. The particles' positions and velocities are initialized randomly within the search space.
During each iteration, the particles update their positions and velocities based on their best-found position, $p_{i}^{\text{best}}(n) = \max_{n} l_{\mathcal{S}}^i(n)$, and the best position found by the swarm up to that iteration, $g^{\text{best}}(n) = \max_{n,i} l_{\mathcal{S}}^i(n)$. The velocity update is given by
\begin{align}
     \delta_{i}(n+1) = w  \delta_{i}(n)& + c_1  r_1 \left[p_{i}^{\text{best}}(n) - l_{i}(n)\right] \nonumber \\ &+ c_2  r_2  \left[g^{\text{best}}(n) - l_{i}(n)\right] ,
\end{align}
where $\delta_{i}(n)$ and $l_{i}(n)$ are the velocity and position of the $i$-th particle at iteration $n$, respectively. Here, $w$ is the inertia weight, $c_1$ and $c_2$ are the cognitive and social coefficients, and $r_1$ and $r_2$ are random numbers uniformly distributed between 0 and 1.
After updating the velocities, the positions of the particles are updated as follows
\begin{align}
l_{i}(n+1) = l_{i}(n) + \delta_{i}(n+1).
\end{align}
Particle positions are constrained to remain within the bounds of the search area $\mathcal{A}$.
After performing $N$ iterations, the global best position, $g^{\text{best}}(N) = \max_{n,i} l_{\mathcal{S}}^i(N)$, is considered the optimal solution found by the algorithm. It is crucial to choose an adequate \ac{numbers of particles $M$ and iterations $N$}  to ensure convergence to the global optimal solution.
\begin{table}
\small
    \centering
        \caption{Simulation Parameters}
    \begin{tabular}{|c|c||c|c|}
    \hline
         Parameter & Value  & Parameter & Value \\
         \hline 
         $\lambda_{\mathcal{R}}$, $\lambda_{\mathcal{C}}$ & 6/km$^2$  & $\lambda_{\mathcal{U}}$   & 10/km$^2$ \\
         \hline
         $\alpha, \beta_0$ & $3,  10^{-6}$  & $w, c1, c2$ & $0.5,~1.5,~2$ \\
         \hline
         $\bar{\gamma}, P_c$ & 5~dB , 0.3  & $P_T$   & 10 ~mW \\
         \hline
         $\sigma^2_{\mathcal{U}}$ & $\sim \mathcal{U}(0.2,0.25)$ & $N_c$   & $\sim \mathcal{U}(50,80)$ \\
     \hline
    \end{tabular}
    \label{table1}
\end{table}
\section{Numerical Results and Analysis}

\label{sec:results}
\begin{figure}[t]
        \centering
        \includegraphics[width=\columnwidth]{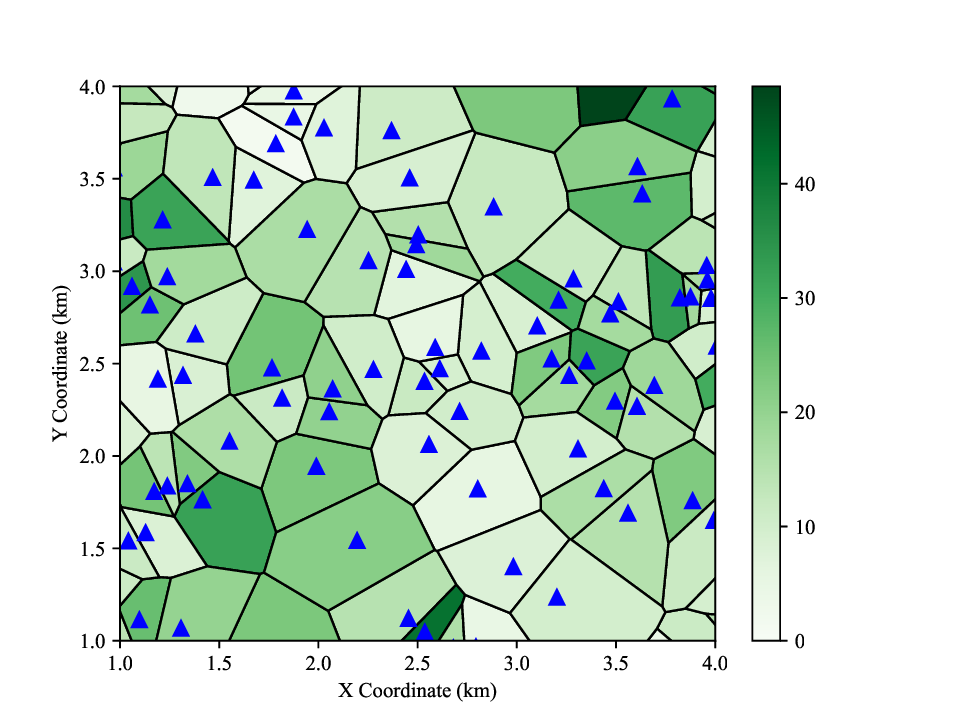}
        \caption{Normalized number of users in each Voronoi.}\label{Voronoi}
\end{figure}

In this section, selected numerical results are provided to verify the effectiveness of the proposed approach for optimizing the location of the SH in F-RAN \ac{settings}. We \ac{distribute the eRRHs and the clusters of the users in} a $5\times5$ km$^2$ area. The simulation parameters are provided in Table \ref{table1}. \ac{Figure}~\ref{Voronoi} depicts the normalized number of users in each Voronoi cell for a realization of the spatial distribution of users and eRRHs within the area over a certain period of time.
To ensure fair performance evaluation, the analysis focuses on a central $3 \times 3$ km$^2$ region (coordinates $1$ km to $4$ km), ensuring uniform interference levels across all eRRHs.

\begin{figure}[t]
        \centering
        \includegraphics[width=\columnwidth]{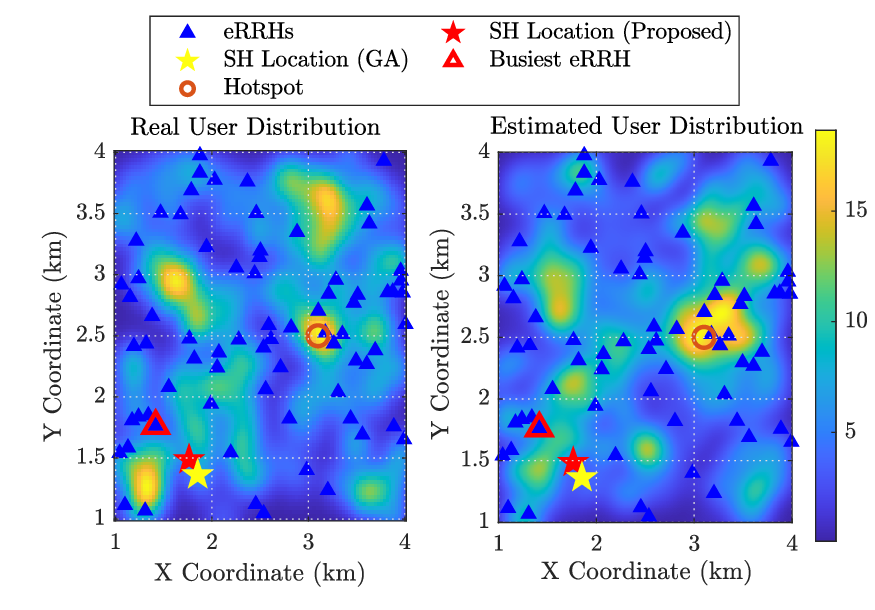}
        \caption{Real user density distribution versus estimated model using the proposed approach.}\label{Coord}
\end{figure}
\begin{figure}[t]
        \centering
        \includegraphics[width=\columnwidth]{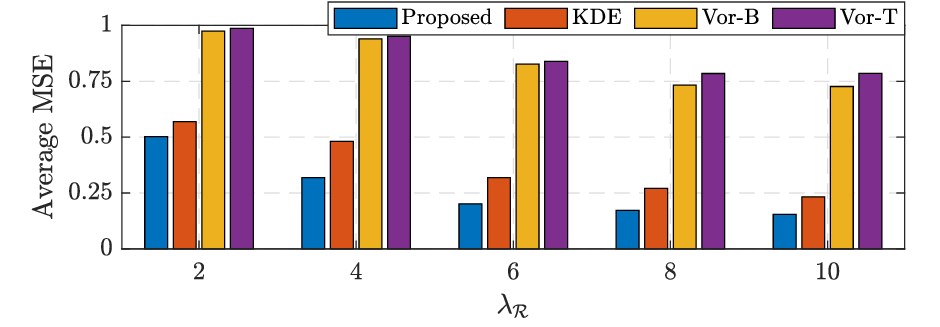}
        \caption{Comparison of MSE values of user density estimation approaches across different values of $\lambda_{\mathcal{R}}$.}\label{UD_compare}
\end{figure}
\begin{figure}[t]
        \centering
        \includegraphics[width=\columnwidth]{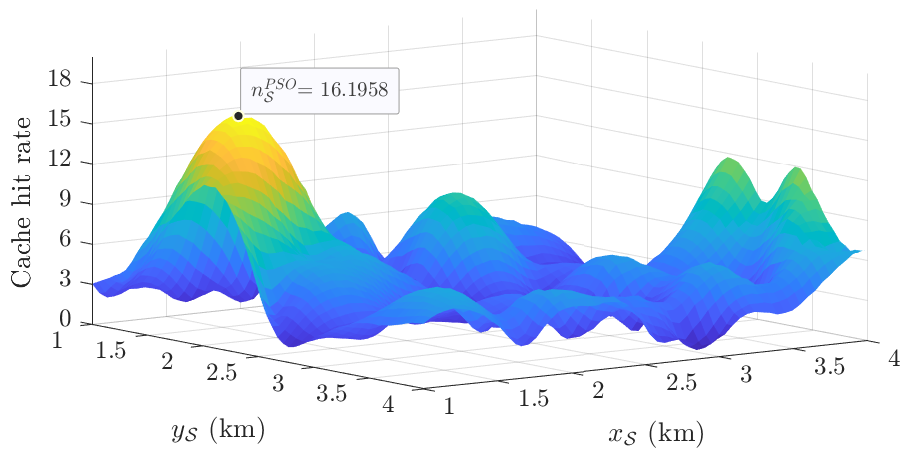}
        \caption{Cache hit rate of the SH in different locations of the map.}\label{3d}
\end{figure}

\begin{figure}[t]
        \centering
        \includegraphics[width=\columnwidth]{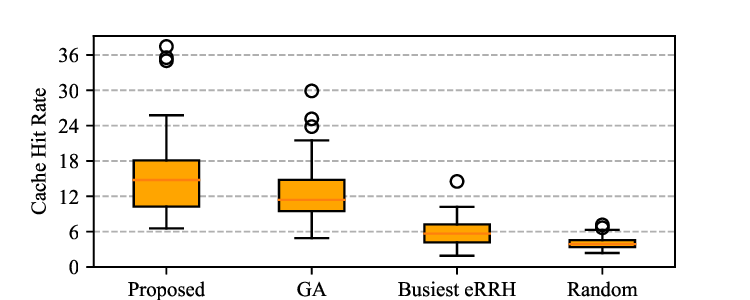}
        \caption{Comparison of the cache hit rate of the SH across various locations.}\label{Comp}
\end{figure}
\ac{Figure}~\ref{Coord} shows the user density distribution and its estimation using our proposed algorithm. As seen, our proposed approach captures the overall dynamics of the user distribution, despite some finer details being missed in the estimated density. The figure also highlights key coordinates for potential SH locations, including the busiest eRRH, the primary user hotspot, and the optimized SH location determined by the proposed and GA algorithms, both using 20 particles/individuals over 20 iterations/generations. 

Figure~\ref{UD_compare} compares the average mean square error (MSE) of user density estimation across the proposed algorithm, kernel density estimation (KDE), and two Voronoi-based approaches: Vor-T, which assigns uniform distributions centered at eRRH locations, and Vor-B, which places uniform distributions centered at the barycenters of each cell's signal dominance profile, as considered in most previous literature~\cite{9647984}. The figure shows the proposed algorithm achieves the lowest MSE and outperforms KDE and the region-based algorithm. While KDE shows moderate accuracy, the region-based method exhibits the highest MSE which indicates its lower effectiveness. Additionally, increasing the number of eRRHs improves estimation accuracy for all methods by providing higher data resolution.

\bc{Figure~\ref{3d} shows the cache hit rate of the SH, $n_{\mathcal{S}}(x_{\mathcal{S}},y_{\mathcal{S}})$, across different locations in the region. The figure highlights that, while there are several local optima where the SH could serve many users, the proposed algorithm effectively achieves the best location with the global optimum hit rate value, where the SH can serve the most users. Moreover, a joint analysis of Figures~\ref{3d} and~\ref{Coord} reveals that the optimal SH location does not essentially coincide with the primary hotspot or the areas with the highest eRRH activity, where the user concentration is the densest. This is because the cache hit rate of SH is influenced not only by user density but also by the spatial distribution of eRRHs and the interference they introduce to the users.}

\ac{Figure}~\ref{Comp} compares of the cache hit rates achieved by the SH across various placement strategies, represented through box plots derived from 50 independent trials. The figure reveals that the proposed approach achieves the highest cache hit rate, both in terms of average and maximum values, outperforming alternative strategies such as the GA approach, placement at the busiest eRRH, and random placement. The proposed approach improves the average cache hit rate by more than 25\% compared to the GA approach, with the average cache hit rate exceeding 50 for the proposed approach while remaining below 40 for the GA approach. While the GA method shows performance closest to the proposed approach, it falls short in terms of both average and peak performance.

\begin{figure}[t!]
        \centering
        \includegraphics[width=\columnwidth]{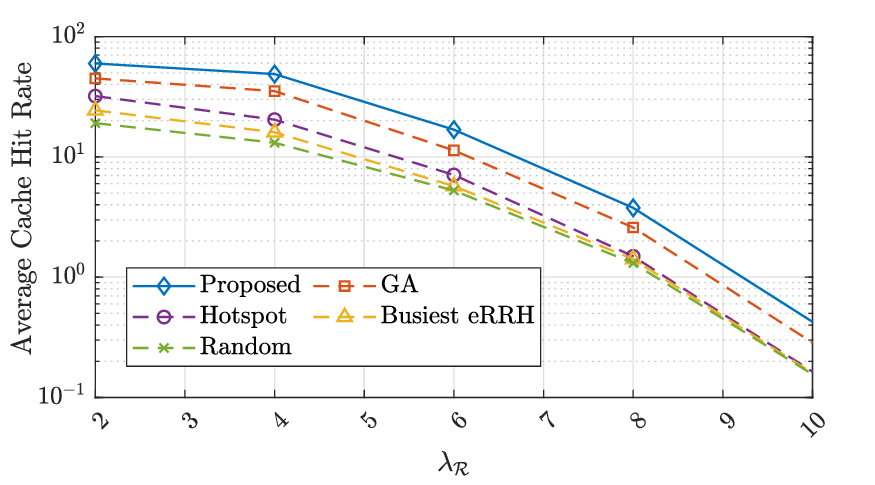}
        \caption{Comparison of cache hit rate of the SH versus $\lambda_\mathcal{R}$ at different locations.}\label{HitRate}
\end{figure}

\ac{Figure}~\ref{HitRate} depicts the cache hit rate of the SH at locations determined by the proposed approach, the GA algorithm, the busiest eRRH, the primary hotspot, and random placements, evaluated across various $\lambda_{\mathcal{R}}$ values. As seen, while the GA-optimized location exhibits better performance compared to the busiest eRRH, the primary hotspot, and random placements, the proposed approach significantly outperforms all alternatives, which demonstrates its superior reliability, scalability, and efficiency over all alternatives.
\section{Conclusion}
\label{sec:conclusion}
\ac{This letter considers optimizing the placement of a SH within an F-RAN architecture}. We introduced a novel approach that leverages the RBF algorithm to estimate user density distribution based on the activity history of the eRRHs. Building on this estimation, \ac{we formulated an optimization problem that minimizes the outage probability and thereby maximizes the cache hit rate across the targeted area via optimizing the SH placement}. \ac{This optimization problem was tackled by employing the PSO algorithm.} Our simulation results validated the effectiveness of the proposed approach, demonstrating its superiority over existing methods and confirming that it can reliably achieve the global optimal solution. Specifically, our proposed algorithm achieves approximately 25\% cache hit rate of improvement compared to the GA approach.
\bibliographystyle{IEEEtran}
\bibliography{main}

\begin{thebibliography}{10}
\providecommand{\url}[1]{#1}
\csname url@samestyle\endcsname
\providecommand{\newblock}{\relax}
\providecommand{\bibinfo}[2]{#2}
\providecommand{\BIBentrySTDinterwordspacing}{\spaceskip=0pt\relax}
\providecommand{\BIBentryALTinterwordstretchfactor}{4}
\providecommand{\BIBentryALTinterwordspacing}{\spaceskip=\fontdimen2\font plus
\BIBentryALTinterwordstretchfactor\fontdimen3\font minus \fontdimen4\font\relax}
\providecommand{\BIBforeignlanguage}[2]{{%
\expandafter\ifx\csname l@#1\endcsname\relax
\typeout{** WARNING: IEEEtran.bst: No hyphenation pattern has been}%
\typeout{** loaded for the language `#1'. Using the pattern for}%
\typeout{** the default language instead.}%
\else
\language=\csname l@#1\endcsname
\fi
#2}}
\providecommand{\BIBdecl}{\relax}
\BIBdecl

\bibitem{7558153}
S.-H. Park, O.~Simeone, and S.~Shamai~Shitz, ``Joint optimization of cloud and edge processing for fog radio access networks,'' \emph{IEEE Trans. Wirel. Commun.}, vol.~15, no.~11, pp. 7621--7632, Aug. 2016.

\bibitem{6928445}
H.~ElSawy, E.~Hossain, and M.-S. Alouini, ``Analytical modeling of mode selection and power control for underlay d2d communication in cellular networks,'' \emph{IEEE Transactions on Communications}, vol.~62, no.~11, pp. 4147--4161, 2014.

\bibitem{D2D}
H.~Xiang, M.~Peng, Y.~Cheng, and H.-H. Chen, ``Joint mode selection and resource allocation for downlink fog radio access networks supported d2d,'' in \emph{2015 11th International Conference on Heterogeneous Networking for Quality, Reliability, Security and Robustness (QSHINE)}, 2015, pp. 177--182.

\bibitem{RW4}
S.~Yan, M.~Peng, M.~A. Abana, and W.~Wang, ``An evolutionary game for user access mode selection in fog radio access networks,'' \emph{IEEE Access}, vol.~5, pp. 2200--2210, Jan. 2017.

\bibitem{RW5}
X.~Wang, S.~Leng, and K.~Yang, ``Social-aware edge caching in fog radio access networks,'' \emph{IEEE Access}, vol.~5, pp. 8492--8501, Apr. 2017.

\bibitem{CERelay1}
C.~Psomas, G.~Zheng, and I.~Krikidis, ``Cooperative wireless edge caching with relay selection,'' in \emph{2017 IEEE International Conference on Communications (ICC)}, 2017, pp. 1--5.

\bibitem{CERelay3}
R.~A. Hassan, A.~M. Mohamed, J.~Tadrous, M.~Nafie, T.~ElBatt, and F.~Digham, ``Dynamic proactive caching in relay networks,'' in \emph{2017 15th International Symposium on Modeling and Optimization in Mobile, Ad Hoc, and Wireless Networks (WiOpt)}, 2017, pp. 1--8.

\bibitem{CERelay2}
J.~Kakar, A.~Alameer~Ahmad, A.~Chaaban, A.~Sezgin, and A.~Paulraj, ``Cache-assisted broadcast-relay wireless networks: A delivery-time cache-memory tradeoff,'' \emph{IEEE Access}, vol.~7, pp. 76\,833--76\,858, 2019.

\bibitem{10818447}
H.~Mokhtarzadeh, M.~Saif, M.~J. Hossain, and J.~Cheng, ``Smart helper-aided f-rans: Improving delay and reducing fronthaul load,'' \emph{IEEE Transactions on Communications}, pp. 1--1, 2024.

\bibitem{7248881}
X.~Chen, Y.~Jin, S.~Qiang, W.~Hu, and K.~Jiang, ``Analyzing and modeling spatio-temporal dependence of cellular traffic at city scale,'' in \emph{2015 IEEE International Conference on Communications (ICC)}, 2015, pp. 3585--3591.

\bibitem{6757900}
D.~Lee, S.~Zhou, X.~Zhong, Z.~Niu, X.~Zhou, and H.~Zhang, ``Spatial modeling of the traffic density in cellular networks,'' \emph{IEEE Wireless Communications}, vol.~21, no.~1, pp. 80--88, 2014.

\bibitem{7083041}
L.~Sardo and J.~Kittler, ``Rbf networks for density estimation,'' in \emph{1996 8th European Signal Processing Conference (EUSIPCO 1996)}, 1996, pp. 1--4.

\bibitem{8108153}
Z.~Zhang, F.~Liu, Z.~Zeng, and W.~Zhao, ``A traffic prediction algorithm based on bayesian spatio-temporal model in cellular network,'' in \emph{2017 International Symposium on Wireless Communication Systems (ISWCS)}, 2017, pp. 43--48.

\bibitem{7733098}
H.~ElSawy, A.~Sultan-Salem, M.-S. Alouini, and M.~Z. Win, ``Modeling and analysis of cellular networks using stochastic geometry: A tutorial,'' \emph{IEEE Communications Surveys \& Tutorials}, vol.~19, no.~1, pp. 167--203, 2017.

\bibitem{9860085}
J.~Han, F.-C. Zheng, J.~Luo, and X.~Zhu, ``Local delay analysis and improvement for user-sbs correlated deployments with random dtx,'' \emph{IEEE Transactions on Vehicular Technology}, vol.~71, no.~12, pp. 13\,007--13\,016, 2022.

\bibitem{8023448}
C.~Saha, M.~Afshang, and H.~S. Dhillon, ``Poisson cluster process: Bridging the gap between ppp and 3gpp hetnet models,'' in \emph{2017 Information Theory and Applications Workshop (ITA)}, 2017, pp. 1--9.

\bibitem{6171996}
H.~S. Dhillon, R.~K. Ganti, F.~Baccelli, and J.~G. Andrews, ``Modeling and analysis of k-tier downlink heterogeneous cellular networks,'' \emph{IEEE Journal on Selected Areas in Communications}, vol.~30, no.~3, pp. 550--560, 2012.

\bibitem{6042301}
J.~G. Andrews, F.~Baccelli, and R.~K. Ganti, ``A tractable approach to coverage and rate in cellular networks,'' \emph{IEEE Transactions on Communications}, vol.~59, no.~11, pp. 3122--3134, 2011.

\bibitem{6524460}
H.~ElSawy, E.~Hossain, and M.~Haenggi, ``Stochastic geometry for modeling, analysis, and design of multi-tier and cognitive cellular wireless networks: A survey,'' \emph{IEEE Communications Surveys \& Tutorials}, vol.~15, no.~3, pp. 996--1019, 2013.

\bibitem{9647984}
F.~Ricciato and A.~Coluccia, ``On the estimation of spatial density from mobile network operator data,'' \emph{IEEE Transactions on Mobile Computing}, vol.~22, no.~6, pp. 3541--3557, 2023.

\end{thebibliography}

\end{document}